\documentclass[twocolumn,aps,prl,superscriptaddress,showpacs,floatfix]{revtex4}
\usepackage{amsmath}
\usepackage{graphicx}

\begin{document}

\title{Light clusters production as a probe to nuclear symmetry energy}
\author{Lie-Wen Chen}
\thanks{On leave from Department of Physics, Shanghai Jiao Tong University,
Shanghai 200030, China}
\affiliation{Cyclotron Institute and Physics Department, Texas A\&M University, College
Station, Texas 77843-3366}
\author{C.M. Ko}
\affiliation{Cyclotron Institute and Physics Department, Texas A\&M University, College
Station, Texas 77843-3366}
\author{Bao-An Li}
\affiliation{Department of Chemistry and Physics, P.O. Box 419, Arkansas State
University, State University, Arkansas 72467-0419}
\date{\today}

\begin{abstract}
Using an isospin-dependent transport model for heavy-ion collisions induced
by neutron-rich nuclei at intermediate energies, we study the production of
light clusters such as deuteron, triton, and $^{3}$He via coalescence of
nucleons. We find that both the yields and energy spectra of these light
clusters are affected significantly by the density dependence of nuclear
symmetry energy, with a stiffer symmetry energy giving a larger yield.
\end{abstract}

\maketitle

The energy per particle of an asymmetric nuclear matter with density $\rho $
and an isospin asymmetry $\delta =(\rho _{n}-\rho _{p})/\rho $, where $\rho
_{n}$ and $\rho _{p}$ are, respectively, its neutron and proton densities,
is usually approximated by a parabolic law \cite{li98}, i.e., 
\begin{equation}
E(\rho ,\delta )=E(\rho ,0)+E_{\mathrm{sym}}(\rho )\delta ^{2},
\end{equation}%
where $E(\rho ,0)$ is the energy per particle of symmetric nuclear matter
and $E_{\mathrm{sym}}(\rho )$ is the nuclear symmetry energy. Although the
nuclear symmetry energy at normal nuclear matter density $\rho _{0}=0.16$ 
\textrm{fm}$^{-3}$ has been determined to be around $30$ \textrm{MeV} from
the empirical liquid-drop mass formula \cite{myers,pomorski}, its values at
other densities are poorly known. Studies based on various theoretical
models also give widely different predictions \cite{ibook}. Lack of this
knowledge has hampered our understanding of both the structure of
radioactive nuclei \cite{oya,brown,hor01,furn02} and many important issues
in nuclear astrophysics \cite{bethe,bom,lat01}, such as the nucleosynthesis
during pre-supernova evolution of massive stars and the properties of
neutron stars \cite{bethe,lat01}. However, recent advance in radioactive
nuclear beam facilities provides the opportunity to study the density
dependence of the nuclear symmetry energy. Theoretical studies have already
shown that in heavy-ion collisions induced by neutron-rich nuclei, the
effect of nuclear symmetry energy can be studied via the pre-equilibrium
neutron/proton ratio \cite{li97}, the isospin fractionation \cite%
{fra1,fra2,xu00,tan01}, the isoscaling in multifragmentation \cite{betty},
the proton differential elliptic flow \cite{lis}, the neutron-proton
differential transverse flow \cite{li00}, the $\pi ^{-}$ to $\pi ^{+}$ ratio %
\cite{li02}, and two-nucleon correlation functions \cite{chen02}.

In this work, we study the production of deuteron, triton, and $^{3}$He in
heavy-ion collisions induced by neutron-rich nuclei by means of the
coalescence model based on the nucleon phase space distribution functions
from an isospin-dependent transport model. It is found that both the
multiplicities and energy spectra of these light clusters are sensitive to
the density dependence of nuclear symmetry energy but not to the
isospin-independent part of nuclear equation of state and the in-medium
nucleon-nucleon cross sections. Therefore, light clusters production in
heavy-ion collisions induced by neutron-rich nuclei provides another
possible method for extracting useful information about the nuclear symmetry
energy.

The coalescence model has been used extensively in describing the production
of light clusters in heavy-ion collisions at both intermediate \cite%
{Gyu83,Aich87,Koch90,Indra00} and high energies \cite{Mattie95,Nagle96}. In
this model, the probability for producing a cluster is determined by its
Wigner phase-space density and the nucleon phase-space distribution at
freeze out. Explicitly, the multiplicity of a $M$-nucleon cluster in a heavy
ion collision is given by \cite{Mattie95} 
\begin{eqnarray}
N_{M} &=&G\int d\mathbf{r}_{i_{1}}d\mathbf{q}_{i_{1}}\cdots d\mathbf{r}%
_{i_{M-1}}d\mathbf{q}_{i_{M-1}}  \notag \\
&\times &\langle \underset{i_{1}>i_{2}>...>i_{M}}{\sum }\rho _{i}^{W}(%
\mathbf{r}_{i_{1}},\mathbf{q}_{i_{1}}\cdots \mathbf{r}_{i_{M-1}},\mathbf{q}%
_{i_{M-1}})\rangle .
\end{eqnarray}%
In the above, $\mathbf{r}_{i_{1}},\cdots ,\mathbf{r}_{i_{M-1}}$ and $\mathbf{%
q}_{i_{1}},\cdots ,\mathbf{q}_{i_{M-1}}$ are, respectively, the $M-1$
relative coordinates and momenta taken at equal time in the $M$-nucleon rest
frame; $\rho _{i}^{W}$ is the Wigner phase-space density of the $M$-nucleon
cluster; and $\langle \cdots \rangle $ denotes event averaging. The
spin-isospin statistical factor for the cluster is given by $G$, and its
value is $3/8$ for deuteron and $1/3$ for triton or $^{3}$He, with the
latter including the possibility of coalescence of a deuteron with another
nucleon to form a triton or $^{3}$He \cite{Polleri99}.

For the deuteron Wigner function, it is obtained from the Hulth\'{e}n wave
function, i.e., 
\begin{equation}
\phi =\sqrt{\frac{\alpha \beta (\alpha +\beta )}{2\pi (\alpha -\beta )^{2}}}%
\frac{e^{-\alpha r}-e^{-\beta r}}{r}
\end{equation}%
with parameters $\alpha =0.23$ \textrm{fm}$^{-1}$ and $\beta =1.61$ \textrm{%
fm}$^{-1}$ to reproduce the measured deuteron root-mean-square radius of $%
1.96$ \textrm{fm} \cite{Ericsson88}. As in Ref. \cite{Mattie95}, we express
the Hulth\'{e}n wave function in terms of $15$ arbitrary Gaussian functions
and determine their strengths and ranges by a least square fit. For triton
and $^{3}$He, their Wigner functions are obtained from the product of three
ground-state wave functions of a spherical harmonic oscillator with its
parameters adjusted to reproduce the measured root-mean-square radii of
triton and $^{3}$He, i.e., $1.61$ \textrm{fm} and $1.74$ \textrm{fm},
respectively \cite{chen86}. Normal Jacobian coordinates for a three-particle
system are then introduced to derive the Wigner functions for triton and $%
^{3}$He as in Ref. \cite{Mattie95}.

The space-time distribution of nucleons at freeze out is obtained from an
isospin-dependent Boltzmann-Uehling-Uhlenbeck (IBUU) transport model (e.g., %
\cite{li97,li00,li02,li96}). For a review of the IBUU model, we refer the
reader to Ref. \cite{li98}. The IBUU model includes explicitly the isospin
degree of freedom through different proton and neutron initial distributions
as well as their different mean-field potentials and two-body collisions in
subsequent dynamic evolutions. For the nucleon-nucleon cross sections, we
use as default the experimental values in free space. In order to study the
effects due to the isospin-dependence of in-medium nucleon-nucleon cross
sections $\sigma _{\mathrm{medium}}$, we also use a parameterization
obtained from the Dirac-Brueckner approach based on the Bonn A potential %
\cite{lgq9394}. For experimental free-space cross sections, the
neutron-proton cross section is about a factor of $3$ larger than the
neutron-neutron or proton-proton cross sections. On the other hand, the
in-medium nucleon-nucleon cross sections used here have smaller magnitudes
and weaker isospin dependence than $\sigma _{\exp }$ but strong density
dependence. For the isoscalar potential, we use as default the Skyrme
potential with an incompressibility $K_{0}=380$ \textrm{MeV}. This potential
has been shown to reproduce the transverse flow data from heavy ion
collisions equally well as a momentum-dependent soft potential with $%
K_{0}=210$ \textrm{MeV} \cite{pan93,zhang93}.

The IBUU model is solved with the test particle method \cite{buu}. Although
the mean-field potential is evaluated with test particles, only collisions
among nucleons in each event are allowed. Light cluster production from
coalescence of nucleons is treated similar as nucleon-nucleon collisions,
i.e., only nucleons in the same event are allowed to coalesce to light
clusters. Results presented in the following are obtained with $20,000$
events using $100$ test particles for a physical nucleon.

For the density dependence of symmetry energy, we adopt the parameterization
used in Ref. \cite{hei00} for studying the properties of neutron stars,
i.e., 
\begin{equation}
E_{\mathrm{sym}}(\rho )=E_{\mathrm{sym}}(\rho _{0})\cdot u^{\gamma },
\label{srho2}
\end{equation}%
where $u\equiv \rho /\rho _{0}$ is the reduced density and $E_{\mathrm{sym}%
}(\rho _{0})=30$ \textrm{MeV} is the symmetry energy at normal nuclear
matter density. In the following, we consider the two cases of $\gamma =0.5$
(soft) and $2$ (stiff) to explore the large range of $E_{\mathrm{sym}}(\rho
) $ predicted by many-body theories \cite{bom}.

\begin{figure}[th]
\includegraphics[scale=1.1]{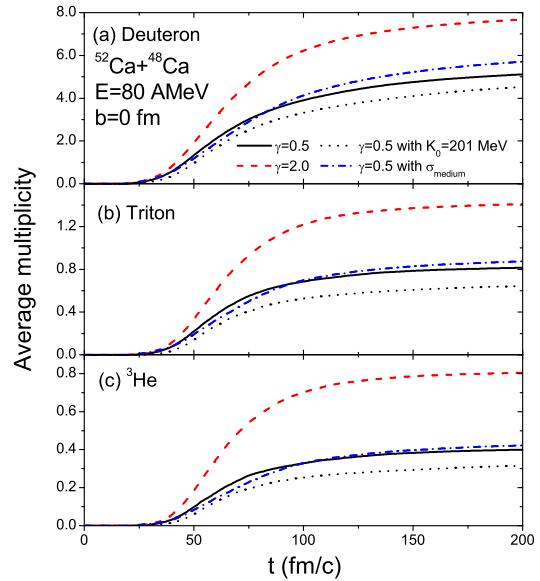} \vspace{-0.5cm}
\caption{{\protect\small Time evolutions of the average multiplicity of (a)
deuteron, (b) triton, and (c) }$^{3}${\protect\small He from central
collisions of }$^{52}${\protect\small Ca+}$^{48}${\protect\small Ca at }$%
E=80 ${\protect\small \ MeV/nucleon by using the soft (solid curves) and
stiff (dashed curves) symmetry energies with a stiff nuclear compressibility 
}$K_{0}=380${\protect\small \ MeV and free nucleon-nucleon cross sections.
Results using the soft symmetry energy and free nucleon-nucleon cross
sections but }$K_{0}=201${\protect\small \ MeV are shown by dotted curves,
while those from the soft symmetry energy and }$K_{0}=380${\protect\small \
MeV but in-medium nucleon-nucleon cross sections are given by dash-dotted
curves.}}
\label{Mult}
\end{figure}

We consider the reaction of $^{52}$Ca + $^{48}$Ca, which has an isospin
asymmetry $\delta =0.2$ and can be studied at future Rare Isotope
Accelerator Facility. In the present study, nucleons are considered as being
frozen out when their local densities are less than $\rho _{0}/8$ and
subsequent interactions do not cause their recapture into regions of higher
density. Shown in Figs. \ref{Mult} (a), (b) and (c) are time evolutions of
the average multiplicity of deuteron, triton, and $^{3}$He from central
collisions of $^{52}$Ca + $^{48}$Ca at $E=80$ MeV/nucleon by using the soft
(solid curve) and stiff (dashed curve) symmetry energies. It is seen that
production of these light clusters from a neutron-rich reaction system is
very sensitive to the density dependence of nuclear symmetry energy. Final
multiplicities of deuteron, triton, and $^{3}$He for the stiff symmetry
energy is larger than those for the soft symmetry energy by $51\%$, $73\%$,
and $100\%$, respectively. This is due to the fact that the stiff symmetry
energy induces a stronger pressure in the reaction system and thus causes an
earlier emission of neutrons and protons than in the case of the soft
symmetry energy \cite{chen03}, leading to a stronger correlations among
nucleons. Furthermore, the soft symmetry energy, which gives a more
repulsive symmetry potential for neutrons and more attractive one for
protons in low density region ($\leq \rho _{0}$) than those from the stiff
symmetry energy, generates a larger phase-space separation between neutrons
and protons at freeze out, and thus a weaker correlations among nucleons.
The larger sensitivity of the multiplicity of $^3$He to the nuclear symmetry
energy than those of triton and deuteron as seen in Fig. \ref{Mult} reflects
the fact that the symmetry energy effect is stronger on lower momentum
protons than neutrons \cite{chen03}.

Although the density dependence of nuclear symmetry energy affects
appreciably the yield of light clusters, changing the incompressibility from 
$K_{0}=380$ to $201$ \textrm{MeV} (dotted curves) or using $\sigma _{\mathrm{%
medium}}$ instead of $\sigma _{\mathrm{exp}}$ (dash-dotted curves) only
leads to a small change in the yield of these clusters as shown in Fig. \ref%
{Mult}. This implies that the relative space-time structure of neutrons and
protons at freeze out is not sensitive to the equation of state (\textrm{EOS}%
) of symmetric nuclear matter and in-medium nucleon-nucleon cross sections.

\begin{figure}[th]
\includegraphics[scale=1.1]{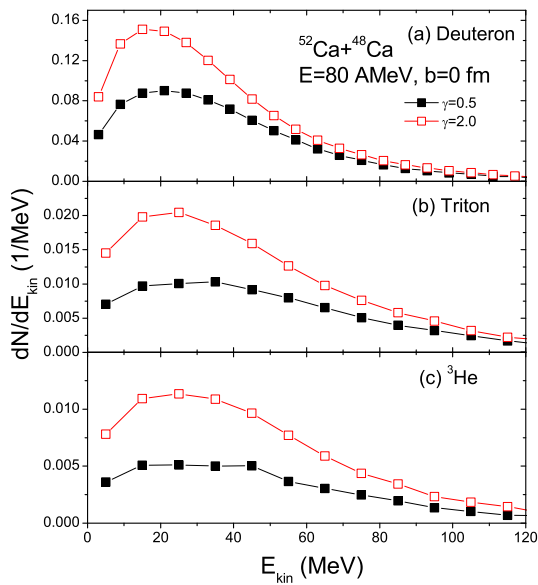} \vspace{-0.5cm}
\caption{{\protect\small Kinetic energy spectra in the center-of-mass system
for (a) deuteron, (b) triton, and (c) }$^{3}${\protect\small He from central
collisions of }$^{52}${\protect\small Ca+}$^{48}${\protect\small Ca at }$%
E=80 ${\protect\small \ MeV/nucleon by using the soft (solid squares) and
stiff (open squares) symmetry energies with a stiff nuclear compressibility }%
$K_{0}=380${\protect\small \ MeV and free nucleon-nucleon cross sections.}}
\label{kinetic}
\end{figure}

The kinetic energy spectra in the center-of-mass system for deuteron,
triton, and $^{3}$He are shown in Fig. \ref{kinetic} for both the soft
(solid squares) and stiff (open squares) symmetry energies. It is seen that
the symmetry energy has a stronger effect on the yield of low energy
clusters than that of high energy ones. For example, the symmetry energy
effect on the yield of deuteron, triton, and $^{3}$He is about $60\%$, $%
100\% $, and $120\%$, respectively, if their kinetic energies are around $10$
\textrm{MeV}, but is about $30\%$, $40\%$, and $85\%$, respectively, if
their kinetic energies are around $100$ \textrm{MeV}. This follows from the
fact that lower energy clusters are emitted later in time when the size of
nucleon emission source is relatively independent of nuclear symmetry
energy, leading thus to a similar probability for nucleons to form light
clusters. Since there are more low energy nucleons for a stiffer symmetry
energy, more light clusters are thus produced. On the other hand, higher
energy nucleons are emitted earlier when the size of emission source is more
sensitive to the symmetry energy, with a smaller size for a stiffer symmetry
energy. The probability for light cluster formation is thus larger for a
stiffer symmetry energy. This effect is, however, reduced by the smaller
number of high energy nucleons if the symmetry energy is stiffer. As a
result, production of high energy light clusters is less sensitive to the
stiffness of symmetry energy. This is different from that seen in the
correlation functions between two nucleons with low relative momentum, where
the symmetry energy effect is larger for nucleon pairs with higher kinetic
energies \cite{chen02}, as they are only affected by the size of the
emission source, not the number of emitted nucleon pairs.

\begin{figure}[th]
\includegraphics[scale=0.9]{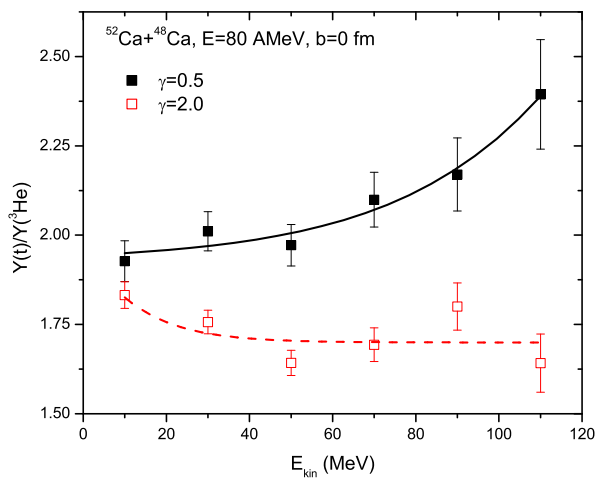} \vspace{-1cm}
\caption{{\protect\small The ratio t/}$^{3}${\protect\small He as a function
of the cluster kinetic energy in the center-of-mass system by using the soft
and stiff symmetry energies. The lines are drawn to guide the eye.}}
\label{RtHe3E}
\end{figure}

The isobaric yield ratio t/$^{3}$He is less model-dependent and also less
affected by other effects, such as the feedback from heavy fragment
evaporation and the feed-down from produced excited triton and $^{3}$He
states. In Fig. \ref{RtHe3E}, we show the t/$^{3}$He ratio with statistical
errors as a function of cluster kinetic energy in the center-of-mass system
for the soft (solid squares) and stiff (open squares) symmetry energies. It
is seen that the ratio t/$^{3}$He obtained with different symmetry energies
exhibits very different energy dependence. While the t/$^{3}$He ratio
increases with kinetic energy for the soft symmetry energy, it decreases
with kinetic energy for the stiff symmetry energy. The symmetry energy thus
affects more strongly the ratio of high kinetic energy triton and $^{3}$He.
For both soft and stiff symmetry energies, the ratio t/$^{3}$He is larger
than the neutron to proton ratio of the whole reaction system, i.e., \textsl{%
N/Z}$=1.5$. This is in agreement with results from both experiments and
statistical model simulations for other reaction systems and incident
energies \cite{Cibor00,Hagel00,Vesel01,Chomaz99}. It is interesting to note
that although the yield of lower energy triton and $^{3}$He is more
sensitive to symmetry energy than higher energy ones, as shown in Fig. \ref%
{kinetic}, their ratio at higher energy is affected more by the symmetry
energy. Moreover, the energy dependence of the ratio t/$^{3}$He is
insensitive to the EOS of symmetric nuclear matter and in-medium
nucleon-nucleon cross sections. These features thus imply that the
pre-equilibrium triton to $^{3}$He ratio is also a sensitive probe to the
density dependence of nuclear symmetry energy.

Isospin effects on cluster production and isotopic ratios in heavy ion
collisions have been previously studied using either the lattice gas model %
\cite{Chomaz99} or a hybrid of IBUU and statistical fragmentation model \cite%
{tan01}. These studies are, however, at lower energies than considered here,
where effects due to multifragmentation as a result of possible gas-liquid
phase transition may play an important role. Except deuterons, both tritons
and $^{3}$He are only about one per event in heavy ion collisions at $80$ 
\textrm{MeV/nucleon}. The number of other clusters such as the alpha
particle is not large either \cite{jacak,borderie}. In this case, the
coalescence model is expected to be a reasonable model for determining the
production of light clusters from heavy ion collisions. Furthermore, the
effect obtained in present study will be enhanced if other clusters are
emitted earlier as the isospin asymmetry of the residue is increased.

In conclusion, using an isospin-dependent transport model together with a
coalescence model for light cluster production, we have found that the
nuclear symmetry energy affects significantly the production of light
clusters in heavy-ion collisions induced by neutron-rich nuclei. More
deuterons, tritons, and $^{3}$He are produced with the stiff nuclear
symmetry energy than the soft nuclear symmetry energy. This effect is
particularly large when these clusters have lower kinetic energies. Also,
the isobaric ratio t/$^{3}$He, especially for higher energy tritons and $^{3}
$He, shows a strong sensitivity to the density dependence of nuclear
symmetry energy. It is further found that light clusters production is not
sensitive to the isospin-independent part of nuclear equation of state and
the in-medium nucleon-nucleon cross sections. The study of light clusters
production in heavy ion collisions induced by neutron-rich nuclei thus
allows us to extract useful information about the density-dependence of
nuclear symmetry energy.

We thank Joe Natowitz for discussions on light clusters production in heavy
ion collisions. This paper is based on the work supported by the U.S.
National Science Foundation under Grant Nos. PHY-0098805 and PHY-0088934 as
well as the Welch Foundation under Grant No. A-1358. LWC is also supported
by the National Natural Science Foundation of China under Grant No. 10105008.

\end{document}